# How to Solve Big Problems: Bespoke Versus Platform Strategies

Atif Ansar

Programme Director, Sustainable Capital-Intensive Industries, Smith School of Enterprise and the Environment, University of Oxford

Senior Fellow, Saïd Business School, University of Oxford

Email: atif.ansar@sbs.ox.ac.uk

Bent Flyvbjerg
First BT Professor and Inaugural Chair of Major Programme Management, University of Oxford's Saïd Business School

Villum Kann Rasmussen Professor and Chair of Major Program Management, IT University of Copenhagen

Senior Research Fellow, St Anne's College, University of Oxford

Email: flyvbjerg@mac.com

**ABSTRACT**
How should government and business solve big problems? In bold leaps or in many smaller moves? We show that bespoke, one-off projects are prone to poorer outcomes than projects built on a repeatable platform. Repeatable projects are cheaper, faster, and scale at lower risk of failure. We compare evidence from 203 space missions at NASA and SpaceX, on cost, speed-to-market, schedule, and scalability. We find that SpaceX's platform strategy was 10X cheaper and 2X faster than NASA's bespoke strategy. Moreover, SpaceX's platform strategy was financially less risky, virtually eliminating cost overruns. Finally, we show that achieving platform repeatability is a strategically diligent process involving experimental learning sequences. Sectors of the economy where governments find it difficult to control spending or timeframes or to realize planned benefits – e.g., health, education, climate, defence – are ripe for a platform rethink.

**ACKNOWLEDGEMENTS**
This work was supported by funding from the Smith School of Enterprise and the Environment, University of Oxford. We thank Ajay Kumar Duraiswamy for his contribution to data collection on NASA's space missions. We are grateful to Ed Hartshorne and Guanpeng Lyu for their research assistance, Madeleine Jones-Casey for proof-reading, and Dora Matok for graphics—their contribution was funded by Foresight Works Ltd. The authors also wish to thank Professor Simon Cowan, Department of Economics, University of Oxford, and Mr Edgar Zapata, NASA for their comments on earlier drafts of the paper.

# INTRODUCTION

The National Programme for Information Technology (NPfIT) was a flagship major programme of the UK Government. At its launch in 2002, NPfIT aspired to digitally transform the National Health Service (NHS) in England by building a number of electronic functionalities: detailed care records for each patient, summary patients records (called Spine), booking of appointments, transfer of prescriptions, ordering and browsing of tests, picture archive and communication system (PACS), clinical decision support system, patient administration system (PAS), plus assorted management tools (Hendy et al., 2004).

By September 2011, the original budget had ballooned from £2.3 billion to £12.4 billion (NAO, 2011; Campion-Awwad et. al., 2014). Following reviews by the National Audit Office, the Public Accounts Committee, and the Major Projects Authority, the UK Government decided to dismantle the NPfIT. The Public Accounts Committee (2013, p. 5) reported, "the benefits to date from the National Programme are extremely disappointing." Even by generous estimates the benefits totalled only one-quarter of the incurred cost of more than £12 billion. The dismal outcome of NPfIT is unfortunately not unique among government major projects: King & Crewe (2013) lament the frequent "blunders of our governments." Ansar et al., (2016) show that even in formidable China at least half the public infrastructure investments are unproductive—that is, net present value negative. Flyvbjerg (2017) has formulated what he calls an "Iron Law" of what is going on: "Over budget, over time, under benefits, over and over again." Why is the typical major project so prone to this law?

Contrast the poor outcomes of NPfIT with another major programme launched in 2001—Apple's iPod, which by 2007 had morphed into the iPhone which Apple CEO Steve Jobs explicitly called "the best iPod we've ever made" (Block, 2007). Since its launch, Apple has sold over 2 billion iPhones and become the most valuable company in the world at a market capitalisation of US$3 trillion. Where NPfIT failed to deliver fewer than 10 apps, the Apple App Store ecosystem has grown from just 3 to over 2.2 million apps, and billions of dollars in revenue.

Why did Apple's iPhone programme succeed where the UK Government's NPfIT failed?

This is *not* a story of private success versus public failure. Both are prone to the Iron Law. Moreover, the private sector was central to the NPfIT: The original inspiration came from Microsoft Chairman Bill Gates in intensive communication with UK Prime Minister Tony Blair. Similarly, private contractors did all major planning and delivery. Equally false would be to assume that a government technology project cannot be compared with private product development: both NPfIT and the iPhone projects used similar financial and human resources over similar periods of time. Different contexts notwithstanding, both projects aimed to scale ambitious services to a large population of end-users. What then explains the very different outcomes of NPfIT versus the iPhone?

We will show that one-off major projects, with a high level of bespoke content, are prone to systematically poorer outcomes when compared with projects built with a repeatable platform strategy. Repeatable projects are positioned to achieve more positive outcomes, faster, cheaper, and at lower risk of failure than one-off

projects.We will further show that achieving platform repeatability is a strategically diligent process involving learning sequences, which are key to explaining the positive outcomes. We propose that governments must radically reallocate resources away from one-off and towards platform strategies in order to meet pressing societal objectives.

One-off projects have theoretical affinity with concepts of "quantum leaps" (Miller and Friesen, 1982), which naïve practitioners often market with superlatives such as "our project is unique…it is the longest / tallest / biggest in the world / first-of-its-kind" (Flyvbjerg and Gardner, 2023). But the truth is that the quantum-leap approach to projects is a recipe for disaster. Examples of this approach include NASA's "grand old style" missions—such as the Space Shuttle (1981-2011) or the doomed Mars Observer (MO) mission launched in 1992. Archetypical megaprojects such as one-off large dams, big science projects like the Manhattan project, or bespoke technology systems such as NPfIT are all instances of aspired quantum leaps. In contrast, repeatable platform projects have theoretical affinity with theories of real options (McGrath, 2004), modular design (Baldwin and Clark, 2000; Flyvbjerg 2021b), and logical incrementalism (Quinn, 1978). Examples of repeatable platforms include computer systems since IBM System/360, containerised shipping, car manufacturing, solar energy, wind energy, and SpaceX.

The paper is organised as follows: The next section reviews contrasting theoretical perspectives from the extant literature on one-off quantum leaps versus repeatable platforms. Then, we describe our methodological approach, followed by empirical results. Empirically, we compare the outcomes of NASA's big one-off projects with SpaceX's platform strategies with the aim of drawing generalisable insights about how best to deliver big, transformative projects at scale. The paper concludes with policy propositions for implementing platform strategies in the real world.

## THEORY

The public sector, even in market economies such as the US or the UK, accounts for nearly half the share of national income (Offer, 2003). Public spending has grown over the past half a century, and so have major public projects. Public spending on long-term commitment goods such as energy, transport infrastructure, healthcare, or defence, typically takes the form of major projects. With growing wealth has come growing demands for government to preserve and promote collective welfare and solve intractable new problems—poverty, crime, economic opportunity, public health, or climate change. A timeless question arises whether public policy responses to big societal problems should occur in big bold leaps or in small repeatable steps.

### BESPOKE STRATEGY

We will interchangeably use "bespoke," "big one-off", and "quantum leap" to describe strategy designed to achieve big goals in one bold step. As a point of departure, Miller and Friesen (1982) offer a useful review of what they call the "Quantum View." "The quantum view advocates that changes in structure must be both concerted and dramatic" (p., 870). Miller and Friesen build the case for rapid and dramatic change by arguing that organizations or societies are greater than the sum of their parts—what they call "gestalts." The presence of such gestalts in their view reduces the capacity to fine tune specific variables one at a time. Such costly piecemeal efforts are either going to lead to no marginal improvement in the system as a whole or become too disjointed and chaotic. Miller (1982) concludes that effective firms will tend to be in one of two states: a common state in which they are undergoing very little or no change or a rarer state in which they will be experiencing dramatic and rapid changes. The quantum view thus postulates that change should occur either in dramatic leaps or not at all. Such an approach will ensure that as little time as possible is spent making turbulent, unsettling, and costly transitions.

Albert O. Hirschman (1967 [2015, p. 1]) wrote long before Miller and Friesen, and he also proposed that societies ought to undertake quantum-leap projects. He called such big plans the "privileged particles of the development process" (p.1) and saw them as "trait making" for society, in contrast to less important "trait taking" projects (p. 118). Hirschman elaborated that "a major development project is a special kind of investment. The term connotes purposefulness, some minimum size, a specific location, the introduction of something qualitatively new." He put forward his argument in favour of big and bold plans with a paradoxical nuance: In launching major policy projects, social planners tended to suffer from a poor conception of the actual cost and effort required to achieve the planned change. Yet, he went on to argue that this *ex-ante* blindness was a good thing. Had planners known the full extent of trouble that lay ahead, they would have never started their big projects. In Hirschman's account, once the major project got underway and found itself in difficulty, planners found reserves of previously undiscovered creativity to overcome the bad surprises and deliver the benefits they set out to do.

Hirschman was well attuned to the pitfalls of quantum leaps (see Flyvbjerg and Sunstein, 2016, pp. 980-982 for a fuller treatment). He identified two explanations— what he called "pseudo-imitation" and "pseudo-comprehensive-program" techniques—for why planners tend to be "providentially ignorant" of risks before embarking on quantum leaps. The twin implication of these pseudo techniques is that planners unconsciously pretend, i.e., delude themselves, "that a project is nothing but a straightforward application of a well-known technique that has been successfully used elsewhere." Moreover, "planners dismiss previous efforts as piecemeal, and portray their own effort as a comprehensive program. With this technique, policymakers give, and are given, the illusion "that the 'experts' have already found all the answers," and all that is needed is faithful implementation" (Hirschman, 1967 [2015] quoted in Flyvbjerg and Sunstein, 2016, p 980). In government major projects, there is thus a tendency for policy units to conceive baroque schemes in isolation and then pressure the civil service to deliver them. The linear thinking of "pseudo-imitation" and "pseudo-comprehensive-program" techniques remains in force today. Deviance from the quantum doctrine is typically deemed a career-limiting heresy. In subsequent sections, when we present evidence of actual outcomes of NASA's quantum leaps, it will become clear that Hirschman was right about the pitfalls he identified for big projects but wrong about the mechanism he thought would save projects from those pitfalls.

Quantum leaps are seductive for their sheer ambition. Such dramatic, concerted jumps readily find societal sublimation: disparate groups hitch their hopes and aspirations to the realization of the supposed "great leap forward" or "a giant leap for mankind." The Hoover dam, or President Kennedy's mission to the Moon are romanticized. Flyvbjerg (2014, p. 8) identifies four such sublimes: technological, political, economic, and aesthetic.

In economics, theories of big—drawing on notions of economies of scale and scope (Stigler, 1958; Silberston, 1972; Chandler, 1990), natural monopoly (Mill, 1848;

Mosca, 2008), or preemptive capacity building (Porter, 1980)—have also advocated for "quantum leaps" in the form of singular, one-off projects, since the mid-nineteenth century. Commitments to largescale are seen as necessary to secure efficient scale and lock competitors out of future rivalry (Ghemawat, 1991; Ghemawat & der Sol, 1998; Hayes & Garvin, 1982, p. 78; Pensrose, [1959] 2009, pp. 89-92; Wernerfelt & Karnani, 1987). Porter (1980, p. 327) recommends, "If economies of scale are large relative to total market demand, an early preemptive capacity move may deny competitors enough residual demand to be efficient. In this case, competitors who invest [later] must invest heavily…or they will have inherently higher costs if they invest on a small scale." Graphically, economies of big are shown as a persistently declining average cost curve as output expands (Ansar & Pohlers, 2014, p 225, Figure 1).

Theories of big have held an enduring sway on corporate strategists and public policymakers. Achieving large scale is deemed necessary for survival and competitive advantage in many industries such as steel (Crompton & Lesourd, 2008), shipping (Cullinane & Khanna, 2000), banking (Cavallo & Rossi, 2001), telecommunications (Majumdar & Venkataraman, 1998; Foreman & Beauvais, 1999), water, or electricity utilities (Kwoka, 2002; Pollitt & Steer, 2011). Mega-mergers are undertaken on the basis of assumed economies of big (Milbourn et al., 2001; Lambrecht, 2004). Similar big moves are witnessed when companies try to pursue new product or geographic markets: Motorola spent US$ 6 billion in technological rollout of its Iridium satellite constellation (Bhidé, 2008); the German steel company ThyssenKrupp splurged over US$ 11 billion to enter the American and Brazilian markets (Ansar, 2012). Even social welfare is deemed to hinge on governments' ability to invest big in long-term "prudential goods" (Offer, 2003, 2006). In the context of economic development, Sachs (2006) has popularized the view that big challenges—such as poverty alleviation, energy and water scarcity, or urbanization can only be solved with "big push" solutions.

**PLATFORMS**
Platforms start small but become big by growing incrementally, i.e., with no big leaps involved. It is not that platforms are less ambitious than quantum leaps. In fact, once a platform matures, it may easily get exponentially bigger than anything thought possible by a one-off quantum leap. Platforms, however, approach their ambition with a series of carefully thought through incremental moves. At each iterative juncture, a platform preserves the option to self-correct, change course (pivot), or abandon and start anew before sunk costs and time become psychologically prohibitive. Platforms are an architected set of parts, subsystems, interfaces, and processes that are shared among a set of applications ("apps")—see, for example, Meyer and Lehnerd, 1997; Cennamo and Santalo, 2013; Zhu and Furr, 2016. Platforms' system architecture is designed to create orderly interactions with less transaction cost among multiple standard and non-standard elements and parties. Consider the example of containerized global shipping. Some of the elements interacting on the global shipping platform are interoperable 20- and 40-foot containers, cranes, vessels, communication satellites, or lighthouses; the parties include shipping lines, port operators, shippers, or regulators among others. Whereas the containers are uniformly standard, vessels—despite many shared modular elements—are not. Yet the protocols of the global shipping platform enable orderly interactions, and shipping at a fraction of the cost pre-container.

Many big technology firms such as Apple, Google, Amazon, and Microsoft are based on platforms. Technology firms such as Airbnb, eBay, and Uber are variants of what is called multi-sided platforms, which have captured the imagination of investors: Airbnb does not own hotel rooms, eBay does not own warehouses, Uber does not own taxis, yet they facilitate interactions among multiple sides (buyers and sellers) at scale. This has led some scholars to take an overly narrow definition of platforms—i.e., capital-light digital systems that make markets (see Hagiu and Wright, 2015 for a discussion). The lineage of platforms, however, belongs to the automotive manufacturing industry (see Steinberg, 2021, for an extensive review of how the terminology and industry applications of 'platforms' emerged from the automobile industry over the course of the 20th century). In this paper we subscribe to the broader definition given in the paragraph above. According to this definition, examples of thoughtfully architected platforms include the automobile industry, and especially electric vehicles; computing hardware; the global shipping industry; the recent space launches, primarily led by SpaceX; modular clean energy generation technologies such as wind or solar; battery storage; electricity transmission systems; and pipelines.

Platforms can be recognized by the following four distinguishing characteristics (see Table 1):

*Repeatability*. Platforms typically start small in a specific niche and undergo an incubation phase of hypothesis testing to discover a formula for replication. Once such a repeatable DNA has been discovered, platforms seek to go viral. Repeatability allows platforms to *scale* in terms of unit volume—that is, supply more of the same thing. Containerized shipping was a niche business for the most part of the late 1950s and 1960s until the Vietnam War created an insatiable demand for the product. Containerized shipping has by now scaled to nearly US$20 trillion of annual trade—a volume traditional shipping could not have supported (Levinson, 2006; *The Economist*, 18 May 2013).

*Extendibility*. Platforms typically start life with a single functionality. With time, the *scope* of available functionality extends exponentially. Computing hardware such as the Apple iPod began life with a single functionality—play 1,000 songs. In fewer than seven years, the iPod had evolved into Apple's iPhone. As of 2021, the Apple App Store supports over 2.2 million apps for iOS. Similarly, Amazon extended from selling books to becoming the "everything store."

*Absorptive capacity*. "Absorptive capacity" refers to an organization's learning processes: the capacity to identify, assimilate, and exploit knowledge from the outside environment and internal knowledge sharing and integration (Cohen and Levinthal, 1989; Zahra and George, 2002; Lane et al. 2006). In typical firms, such learning processes can become disjointed or myopic (Levinthal and March, 1993; Lane et al. 2006). In platforms learning is codified in "design rules [that] establish strict partitions of knowledge and effort at the outset of a design process. They are not just guidelines or recommendations: they must be rigorously obeyed in all phases of design and production" (Baldwin and Clark, 2000, p. 6). Operationally, the absorptive capacity enabled by the design rules allows for multiple streams of research and development (R&D) to consolidate in evermore complex wholes. The image of the video game of Tetris and how variously shaped blocks combine to create a platform may help to form a mental picture. The precise nature of subunits (or modules) and the timing by which they might become absorbed into a growing platform need not be preordained. Apple's iPhone was a serendipitous result of multiple experiments—some successful such as the iPod's solid state hard drive, the iPod Touch's touch screen, or the launch of iTunes and some failed such as Apple's tie-up with Motorola to produce a joint-venture phone. The failure provided Apple with the crucial outside knowledge to integrate the multiple streams into a powerful new platform.

*Adaptive capacity*. Platforms start simple and evolve into complex adaptive systems (see Biedenbach and Müller, 2012). Platforms can be repurposed and upgraded. The rigid partitions and interface design rules of a platform, paradoxically, unleash tremendous flexibility. Computer hardware, starting with the case of IBM System/360—the first modular mainframe computer—is a compelling case study to which we will return in the results section. Any user of a computer, e.g., a PC-laptop, knows that computers can be easily configured—a different computer for a different user using the same core building blocks. Modules can be added or removed without redoing the entire machine. Peripherals, such as printers or scanners, enable more specialized functions on command, making possible an unprecedented mixing and matching at will. The adaptive system architecture of a computer allows for compatibility across systems. Circuits designed for one system can be plugged into another, removing costly bespoke parts across systems. Similarly, individual components innovate along independent trajectories; new iterations of components—such as a new graphics card—can be added piece-meal to upgrade a machine incrementally without discarding the whole machine each time. Every couple of years, a next generation model is released to the market—in effect overcoming technological obsolescence. The various modules interact: and are further developed over time, allowing computers to go from being able to do a few things poorly to do a lot of things really well.

Platforms differ from quantum leaps in two primary assumptions about the world. Quantum leapers believe, or pretend to believe:1) The world is a linear, controlled, environment (a la Hirschman's "pseudo-imitation" technique); 2). Experts possess all the necessary information and skills to marshal the risks that may arise in the future (a la Hirschman's "pseudo-comprehensive-program" technique). Platformers, in contrast, believe: 1) The world is a dynamic, complex adaptive system (Baldwin and Clark, 2000); 2). Future adjustments to risk in the external environment cannot be pre-ordained. Instead, carefully thought through options must be designed to flexibly respond to multiple probable worlds in the future (see McGrath, 1997; 2004 on "real options").

**TABLE 1: PLATFORM QUALIA**

| Platform Qualia | Description and Example |
|---|---|
| *Repeatability* | Platforms typically start small in a specific niche and undergo an incubation phase of hypothesis testing to discover a formula for replication. Once such a repeatable DNA has been discovered, platforms go viral. For example, global shipping containers. |
| *Extendibility* | Platforms typically start life with a single functionality. With time, the *scope* of available functionality extends exponentially. For example, Apple iPod began life with a single functionality—play music. Its descendant, Apple iPhone, supports 2.22 million apps. |
| *Absorptive capacity* | Platforms absorb and integrate multiple streams of research & development (R&D)—from internal and external sources of knowledge—to form evermore complex wholes. For example, Apple iPhone was the absorptive result of iPod solid state hard drive, Apple Touch's touchscreen and a failed joint-venture with Motorola. |
| *Adaptive capacity* | Platforms start simple and evolve into complex adaptive systems. Platforms can be repurposed and upgraded. For example, computer hardware. |

Platformers' beliefs about the world and uncertainty translate into a series of operational design choices at the sub-component level. Quantum leaps are typically indivisible: one either has a dam or no dam—there is no sense in talking of a 90% complete dam across a ravine, or a 90% completed bridge. Ansar et al. (2017) liken big one-offs to Humpty Dumpty—if they come apart, they cannot be put together again. Platforms are consciously de- and recomposable. They can be disassembled and reassembled at discretion into different combinations. Eight tools enable the scaling-up or down of platforms: 1) splitting; 2) swapping; 3) augmenting; 4) modernizing; 5) miniaturizing; 6) excluding; 7) inverting; and 8) porting (see Baldwin and Clark, 2000, Ch. 5)[1].

To sum up, the outcomes of platforms are unmissable: markets with platforms make services faster, better, cheaper, and more omnipresent. These forces, undoubtedly, cause disruption to the status quo—platforms are not popular among those who find it hard to keep pace. When a platform player emerges in an industry, evidence to watch out for is as follows: 1) Industry cost curves come down drastically; 2) Speed of service increases; 3) Scale of the market served goes up; 4) Scope of the available services increases; 5) The variety of parties served goes up.

---

[1] Through *splitting*, hardware modules are decomposed further to achieve intensive functional specialization (e.g., the shipping container split breakbulk cargo into unitized cargo). Through *swapping*, multiple peripheral components are made to work with the same core modules (e.g., the same USB ports can support a printer or a smartphone). *Augmenting* supplies new modules that perform previously undiscovered functions (e.g., Wi-Fi receivers in previously unconnected devices). Through *modernizing*, iterative improvements at the component level or generational improvements at the system level—the technological cycle—are achieved (e.g., successive iPhones). Through *miniaturization* the physical dimensions or mass of individual components is reduced (e.g., microchips). Through *excluding*, a module is discarded without discarding the system (e.g., punch cards were discarded). Through *inverting*, redundant functions are consolidated into single new modules (e.g., DVDs could not be played in CD players but now both can be accommodated). Finally, *porting* allows modules to be used (or reused/recycled) across systems and across time, which enables compatibility (e.g., Intel chips can work across many computers). Moreover, through porting, distinct systems are linked through common components enabling networks (e.g., fiber-optic cables link up many computers into the Internet).

Going by the eye-popping valuations of platform companies such as Apple, Google, Amazon, or Microsoft, the practitioner's grasp of the power of platforms perhaps exceeds that of the professional economist. Albert O. Hirschman in his attempt to encourage quantum leaps had expressed "the expectation that a sequence of further development moves will be set in motion" (Hirschman, 1967 [2015, p. 1]). It turns out that platforms, by starting small and thinking big, come closer to Hirschman's expectation than singular big bold leaps.

## METHOD

In this section we present the methods by which the empirical analyses were conducted before presenting their results. For reasons of economy, we focus on one industry, space exploration. The space industry offers an unusual but useful empirical setting for the following reasons: Exploring and populating outer space is a big problem. Since its founding in 1958, NASA has tackled this problem through big one-off missions, described by Donna Shirley, a manager on NASA's Pathfinder mission, as "magnificent mission[s] in the grand old style". Studying space allows us to gain deep insight into how organizations go about solving big problems with big projects.

In comparison to NASA, SpaceX deployed a disciplined platform strategy from 2002. Few industries, if any, exhibit such a clear, near-binary, divergence of strategic orientations. Building on Seawright and Gerring (2008, p. 304), the selection of space as an industry-level case-study follows a "most similar/most different" research design. In such a research design the cases compared against each other are alike in many (ideally *all*) aspects except for the theoretical variable of interest. NASA's space missions share many similarities with SpaceX's in their goal (bringing payload into outer space), technological and managerial complexity, unforgiving constraints placed by laws of physics, and risk of failure. The key difference between the two is that NASA followed a bespoke strategy (used interchangeably with quantum leap or big one-off strategy) and SpaceX followed a platform strategy. Both strategies were deliberate and considered. NASA and SpaceX thus stand proxy for the theoretical variable of interest: does a bespoke or a platform strategy outperform in solving a big problem? Maximum variation cases, in representing the full variation of the population, are powerful tools for generalizability (Flyvbjerg, 2006; Seawright and Gerring, 2008).

A final reason for our choice of the space industry is data access. Unusually for organizations that do big projects, NASA and SpaceX have made large amounts of data available to researchers. NASA is a public sector entity and thus subject to extensive scrutiny. To its credit, NASA has made data access to convenient and transparent, where common practice among sponsors of big projects is to hide or obfuscate data (Ansar, 2017, Ansar et al., 2016; Flyvbjerg, 2014). Duraiswamy (2017, p 26, Table 3.2) collected NASA data from NASA budget disclosures, U.S. government audit reports, U.S. congressional hearings, logs of space launches, industry data, company disclosures, academic articles, news articles, and documentary programs (Congressional Budget Office, CBO, 2005; Emmons et al., 2007; Butts and Linton, 2009; Government Accountability Office, GAO, 2009, 2010, 2011, 2012, 2013, 2014, 2015, 2016, 2017; Office of the Inspector General, 2012). We emphasized written sources because we know from previous research that they are more reliable than interviews and other oral sources (Glueck and Willis, 1979; Ansar et al. 2016).

Separately, we collected data on SpaceX missions to enable the comparative approach outlined above. SpaceX is a privately held company and accordingly does not make regulatory filings. Even so, since many of the SpaceX launches were funded by NASA, we were able to piece together data on SpaceX based on NASA data. Moreover, SpaceX itself is relatively forthcoming with data about its space missions. Its story is well-documented in the secondary literature and by space historians and journalists, which also helped us in collecting the data for study. We rigorously quality controlled the data, verifying that all included data are valid and reliable.

At the level of industry, we carry out a case study (*N* = 1), nested within which is a comparative case analysis of two organizations (*N* = 2), nested within which is a study of a large number of space missions, (*N* = 203), of which 181 were planned and managed by NASA (1963-2019) and 22 by SpaceX (2002-2021). The SpaceX sample is smaller than the NASA sample for the simple reason that SpaceX has flown many fewer missions than NASA because SpaceX is decades younger. The research design is exploratory and was chosen to spur research in the field rather than be the final word on the matter. The paper concludes with four formalised policy propositions, which we hope will enable future research of testing and refininement.

In terms of performance measures, we compare cost, cost overrun, speed-to-market, schedule overrun, and scalability of NASA vs SpaceX missions. For cost performance, we analyze both absolute cost, e.g., payload measured as cost per kilogram to Low-Earth Orbit (LEO), and cost overrun measured as the actual outturn costs expressed as a ratio of estimated costs (see Ansar et al., 2016 for further discussion). Speed-to-market is measured as the actual delivery time from the date of the final investment decision (FID) or contractual close to the date of first successful launch. Our speed-to-market measure excludes pre-planning prior to FID. Schedule overrun is measured as actual duration expressed as a ratio of estimated duration costs (see Ansar et al., 2016 for details). Finally, we measure scalability as number of launches per year.

Two caveats apply. First, when measuring cost overrun, the baseline for NASA is the cost estimate available at the time of the final investment decision (FID), whereas the baseline for SpaceX is the estimated procurement of contract cost to the end-customer such as NASA or the U.S. Department of Defense —i.e. SpaceX's bid amount at the time of tender award. Second, SpaceX total costs exclude any contract management costs incurred by the end-customer. The latter caveat is considered a minor issue, especially for comparisons of percentage cost overruns (also see Zapata, 2017a, 2017b). The first caveat is more serious, because as we point out in Flyvbjerg et al. (2018), different baselines for calculating overrun -- here FID for NASA and contract award for SpaceX -- could result in comparing apples and oranges. In an ideal world, we would have available SpaceX's internal cost estimates at final investment decision and a comparable tally of actual outturn costs at successful mission launch. However, as a privately-held company SpaceX does not publish cash flow statements, its annual capital expenditure, or how that capex is apportioned across missions. What researchers can access is the bid amount at contract award and the final outturn cost borne by SpaceX customers, who then make such data available in the public record. This creates a methodological risk that SpaceX has been grossly underpricing its bids to win contracts. Typically, uncertainty goes down from FID to contract award (Cantarelli et al., 2012) as the FID budget is based on internal cost estimates, whereas the cost estimates at contract award price in market's expectation of the total cost.

By using the available numbers, we are therefore at risk of giving SpaceX an unfair advantage over NASA in measuring costs and cost overruns. To assess this risk, we carried out sensitivity analyses, which show that the difference between NASA and SpaceX is much too large to have been caused by differences in how costs are measured. First, as the results sections show, SpaceX's absolute costs are nearly 10 times lower than NASA's and SpaceX's cost overruns are 80 times lower than NASA's. Second, data about SpaceX's funding rounds and revenues are publicly available. We used these data to test the sensitivity of whether SpaceX could likely underprice its contracts 10 times and remain solvent. We verified that SpaceX is not grossly underpricing its contracts. Finally, NASA has independently verified SpaceX's internal development costs giving further credence that SpaceX contract values are representative of its internal costs (see Zapata, 2017a, 2017b, 2018). What we measure is therefore a real difference between NASA and SpaceX and not a difference caused by differences in accounting. When interpreting the results below it should be kept in mind, however, that they are likely somewhat biased in favour of SpaceX due to the difference in baselines.

**RESULTS**
"The space race is dominated by new contenders," claims *The Economist* (18 October 2018). Chief among them is SpaceX that has gone from a ridiculed upstart in 2002, when Elon Musk founded it, to capturing nearly two-thirds of the global commercial launch market by 2018. SpaceX deployed a disciplined platform strategy to outperform its competitors, public and private.

We report comparative results—NASA vs SpaceX—on four variables (cost, speed-to-market and schedule, and scalability), measured for the sample described above. Across these metrics, SpaceX's platform strategy has vastly outperformed NASA's bespoke strategy in some cases by multiple orders of magnitude.

Before reporting the results in detail, it is worth reiterating with emphasis that the aim is not to put down the public sector and aggrandize private enterprise. The aim is to illustrate that platform-based projects, whether adopted by public or private enterprise, systematically outperform quantum-leap projects. Moreover, the success of SpaceX is the success of NASA because NASA is the largest user and hence beneficiary of SpaceX's platform strategy, together with the U.S. Department of Defence. Conversely, SpaceX owes its very existence to contracts and revenues from NASA. Although SpaceX has executed a platform strategy with skill since 2002, it was NASA, frustrated by the underperformance of its own projects, provided the impetus toward a platform strategy from spaceflight in the mid-2000s, by rejecting the conventional approach "using cost-plus contracts (in which NASA shouldered all the economic risk of investing in space) to fixed-price contracts (in which risk was distributed between NASA and their contractors)" (Weinzierl and Sarang, 2021, p. 6; also see Zapata, 2017a on NASA's COTS/CRS Program[2]). The rise of SpaceX can be seen as a natural experiment pitting the NASA of before the mid-2000's against the NASA after 2005 when it began to seriously pursue a platform strategy, via SpaceX in addition to its traditional approach.

*Cost performance*
With respect to cost performance, we make the following observations:

Using cost per kilogram to Low-Earth Orbit (LEO)—a standard industry metric to compare costs across space systems-Figure 1 shows that SpaceX's rockets have dramatically reduced costs to orbit. SpaceX Falcon Heavy's cost of US$1,400 per kg is 700 times cheaper than Vanguard—the first family of NASA's rockets—44 times cheaper than the retired Space Shuttle programme and 4 times cheaper than Saturn V—the rocket that took humans to the Moon in 1969 on the Apollo 11 mission.

But SpaceX rockets are not only competitive when compared with historic flights. They are also competitive for present-day flights. Thus, prices for payload on a Falcon Heavy launch start as low as US$90 million—about 5 times cheaper than the Delta IV Heavy made by United Launch Alliance (ULA)—jointly owned by Boeing and Lockheed Martin. And less than a third of the price of its closest competitor, the Russian Proton family of rockets, which has been in service since the 1960s.

**Figure 1: Launch Cost Per Kilogram to Low-Earth Orbit (LEO) (US$ Thousands)**

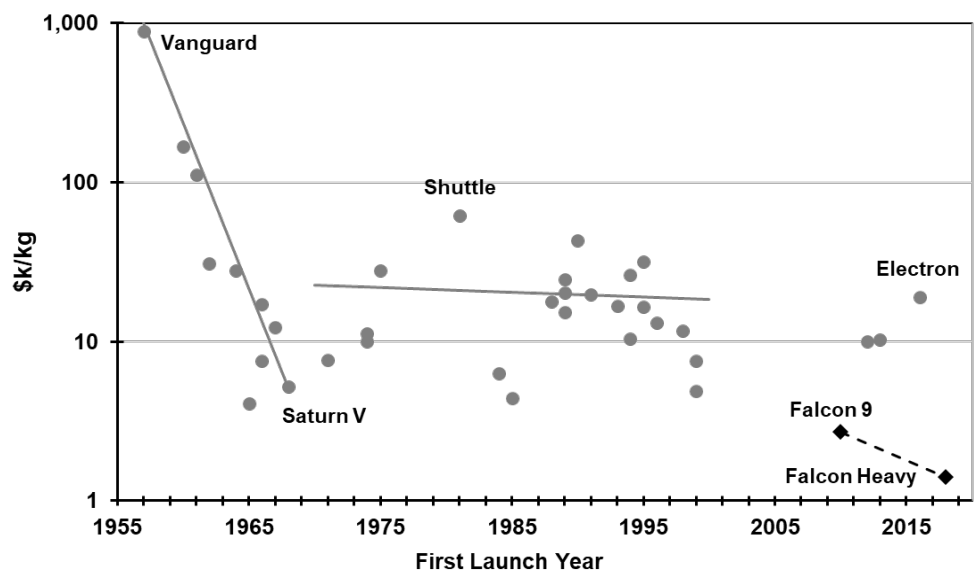

Sources: Jones (2018) and Zapata (2017b)

SpaceX also outperforms in terms of development cost. NASA (2010) verified SpaceX's total development cost for the Falcon 9 rocket at approximately US$390

---
[2] Commercial Orbital Transportation Services (COTS) and Commercial Resupply Services (CRS).

million (including US$90 million towards the Falcon 1 rocket) in year-of-expenditure prices 2002-2009. NASA's 2010 report goes on to estimate "the predicted cost to develop the Falcon 9 if done by NASA would have been between $1.7 billion and $4.0 billion." SpaceX's Falcon 9 system was thus 4-10 times cheaper in development cost than NASA's own estimate.

Similarly, for heavy-lift rockets, SpaceX's Falcon Heavy far outperforms rivals. For the Apollo programme (which included 15 flight-capable Saturn V's rockets) the US spent US$25.4 billion in year-of-expenditure prices (1961-1972) or approximately US$150-160 billion in today's prices (United States Congress, 1973 p. 1271). The development cost of the Saturn V rockets accounted for about 40% of the total spend on the programme or roughly US$60 billion today. NASA's contemporary heavy-lift rocket, the Space Launch System (SLS) has a cost over US$21.2 billion in year-of-expenditures dollars 2011-2021. In comparison, SpaceX's Falcon Heavy, with seven total landings under its belt as of January 2022, has a development cost between US$500-750 million (*CNBC* 8 February 2018). Berger (2018) correctly observes, "The Falcon Heavy is an absurdly low-cost heavy lift rocket."

SpaceX's platform strategy also outperforms NASA in terms of cost adherence, measured as cost overrun on the final approved baseline budget. Figures 2a & 2b and Table 2 show cost overruns of NASA versus SpaceX's missions.

- In terms of frequency of cost overrun: Of the 181 NASA missions in our refence class, we had cost overrun data for 118 missions: 9 in 10 suffered a cost overrun. For SpaceX, the comparable number is 5 in 10, which is exactly what a good portfolio manager would aim for.
- In terms of magnitude of cost overrun: NASA's actual costs were on average +90.0% higher than estimated costs with a median of +45.8% indicating that the distribution of cost overrun has a heavy skew to the right (i.e., going over budget). A Mann-Whitney U test of overall cost neutrality provided conclusive evidence that NASA's budgets were systematically biased towards underestimation (p = 0.004). Thus, there is a strong bias towards adverse outcomes with NASA, as shown in Figure 2a. For SpaceX the comparable numbers are an average cost overrun of +1.1% and a median of +1.5%, indicating little skew, and in the opposite direction. Unlike NASA, SpaceX cost under/overruns are tightly distributed around a mean forecasting error that approximates zero.
- Finally, we tested whether the NASA distribution of cost overruns is different form the SpaceX distributions. Using a Wilcoxon test (p<0.0001), we found overwhelming evidence that SpaceX outperform NASA. SpaceX distribution of cost overruns is far more attractive (lower average than median, much fewer extreme values) than the NASA distribution.

**Table 2: Space missions – NASA vs SpaceX Cost Overruns**

| Entity | Number of cases (N) | Average cost overrun (%) | Standard Deviation | Median cost overrun (%) | Frequency of projects cost overrun |
|---|---|---|---|---|---|
| NASA | 118 | 90.0 | 100.3 | 45.8 | 9 in 10 |
| SpaceX | 16 | 1.1 | 10.0 | 1.5 | 5 in10 |

Source: Authors' Database

**Figure 2a: Density Trace and Mean (Vertical Lines) of Costs Overruns: NASA ($N$ = 118) vs SpaceX ($N$ = 16)**

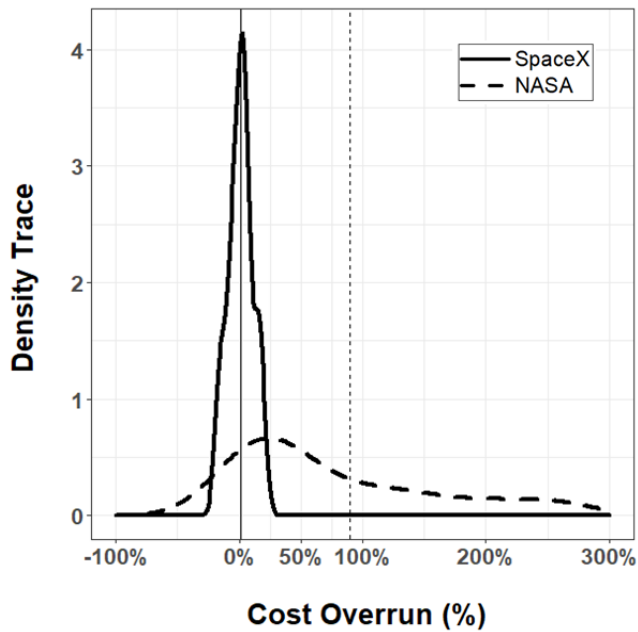

Source: Authors' Database

**Figure 2b: Costs Overruns Over Time: NASA vs SpaceX**

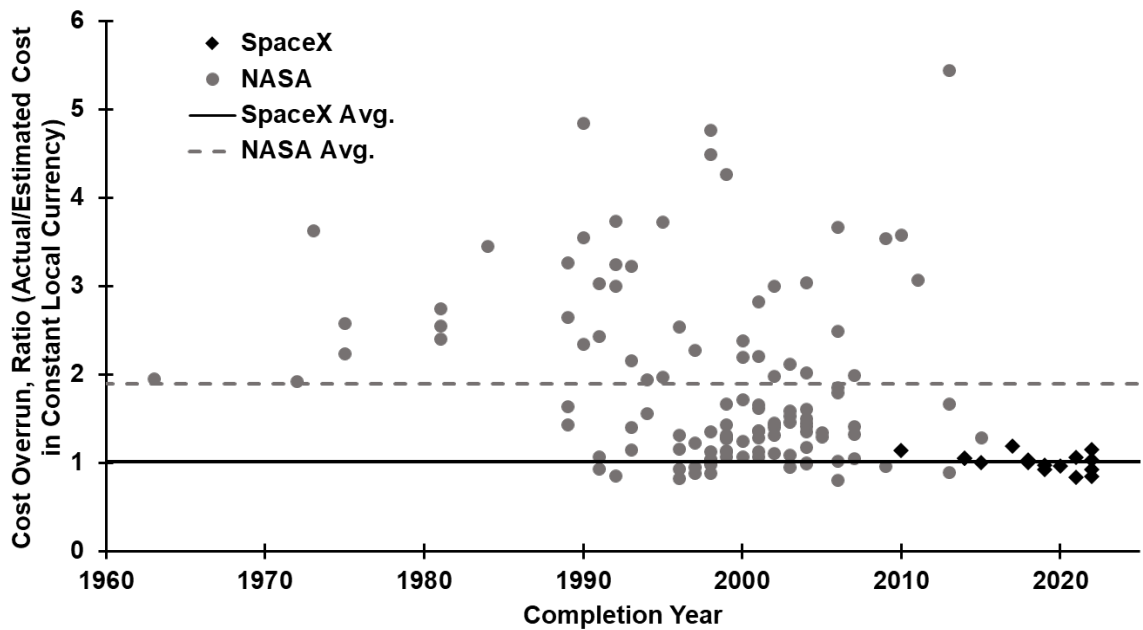

Source: Authors' Database

*Speed-to-market and iterations*
We make the following observations:

In terms of speed-to-market and iterative updates, SpaceX is on average nearly twice as fast as NASA. In terms of actual delivery time, SpaceX projects have a shorter actual duration from the decision to build to completion. On average SpaceX projects took 49.2 months (~4 years) whereas NASA's took 82.3 months (~7 years). This difference is statistically significant (non-parametric Wilcoxon test, $p < 0.0001$). As its platform has stabilized, SpaceX's speed-to-market has been persistently trending lower towards 2 years (see Figure 3a), whereas NASA's is trending back upwards (see Figure 3b). Since our estimates of actual delivery times exclude pre-planning prior to final investment decision or contractual close—SpaceX's relative speed is in fact faster than our deliberately conservative calculation.

**Figure 3a: SpaceX's Speed-to-Market**

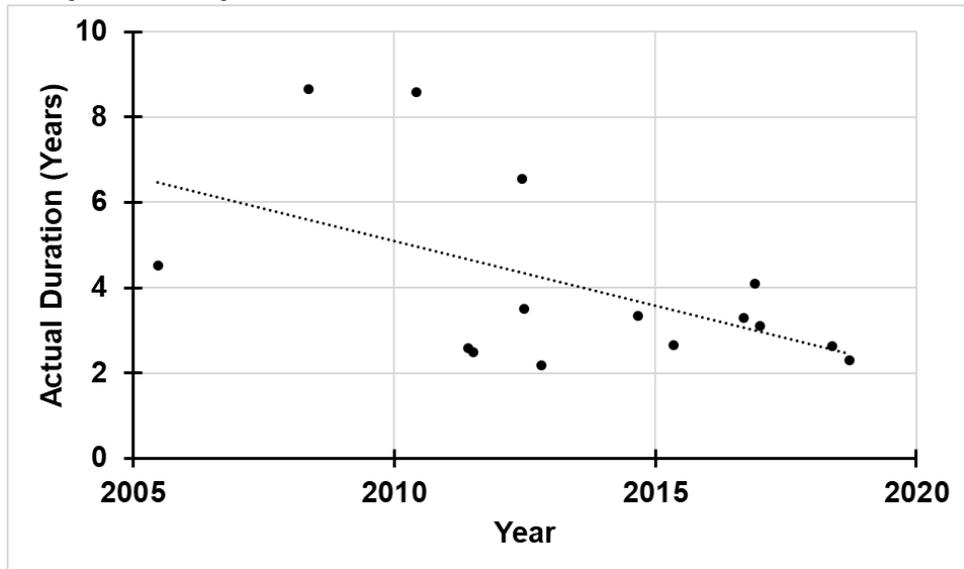

Source: Authors' Database

**Figure 3b: NASA's Speed-to-Market**

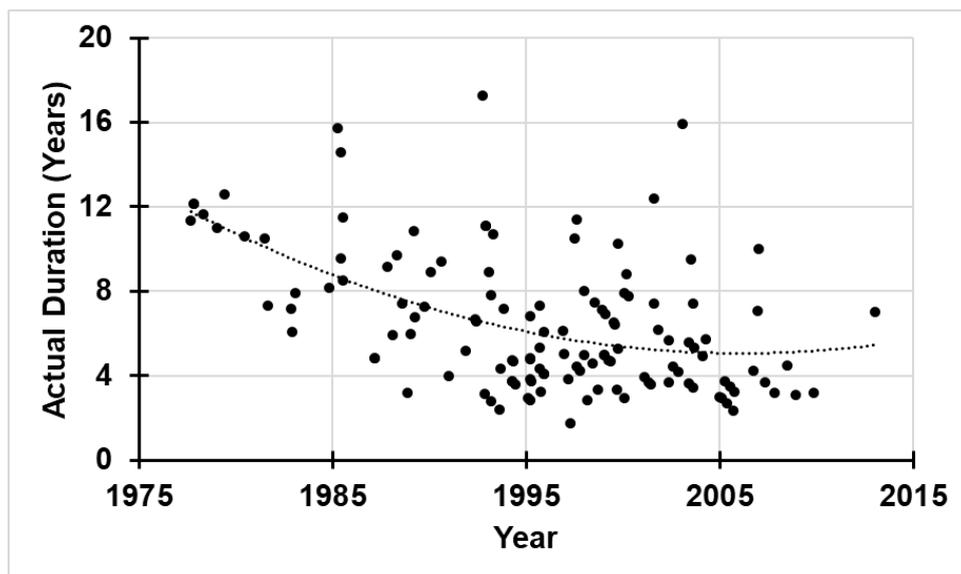

Source: Authors' Database

SpaceX's speedy achievements since its founding in 2002 cannot be understated. SpaceX has thus far built the following systems: Falcon 1, Falcon 9, Dragon, Falcon Heavy, and the Starship (previously named the Starhopper). The essence of a rocket is the engine. SpaceX built its first Merlin engine within 3 years, with its fair share of failures. In the past it has taken at least 7 years to build a rocket engine (Berger, 2021). At the time of writing, the complete line-up of SpaceX rocket engines includes Merlin, Kestrel, and Raptor for use in launch vehicles, Draco for the reaction control system of the Dragon series of spacecraft and SuperDraco for abort capability in human crew missions of the Dragon. Each engine has undergone rapid iterative updates: the Merlin family alone includes 1A, 1B, 1C, 1C Vacuum, 1D, 1D Vacuum.

For SpaceX launch vehicles, the Falcon 1 had its first successful launch in March 2006 (within four years of SpaceX's founding). By November 2008, the next-generation Falcon 9 had it first full-duration test-firing (2 years). Falcon 1 was retired in July 2009 after its fifth and final launch. Much like newer versions of the iPhone, the Falcon 9 was upgraded to Falcon 9 v1.1 in 2013, Falcon 9 Full Thrust in 2015, and finally to Falcon 9 Block 5 in 2018. The first stage of Falcon 9 is designed to retropropulsively land, be recovered, and reflown. If Falcon 9 is the SpaceX's iPhone, the Falcon Heavy is the iPad—a much larger rocket for heavy lifts into space. With the ability to lift approximately 64 metric tons into lower-Earth orbit, Falcon Heavy can lift more than twice the payload of the next closest operational vehicle, the Delta IV Heavy. Falcon Heavy's first launch took place in February 2018. The Falcon 9 and Falcon Heavy have been launched 137 times (including one in-flight failure) and reflown 78 times more than the NASA Space Shuttle in its thirty-year history from 1981-2011.

Even as the Falcon family of rockets were evolving, SpaceX developed the Dragon spacecraft, which was in service by December 2010, when SpaceX became the first private company to successfully launch, orbit, and recover a spacecraft. By 2012, the Dragon become the first privately built spacecraft to deliver cargo to the International Space Station. In May 2020, SpaceX flew a human crew aboard its Dragon spacecraft. Later in 2020, the first full-sized Starship prototype made a five-hundred-foot test flight. SpaceX achieved the first successful suborbital flight and landing of a full Starship prototype by May 2021. SpaceX's pace in bringing new systems and iterative upgrades to the market has been nothing less than stunning compared with NASA.

Take NASA's Mars Observer mission as a case in point. With a 17-year planning and development cycle and a cost of over US$1.3 billion in 2000 prices, the Mars Observer was slow to market and costly. The Mars Observer was launched in September 1992. On August 21, 1993, three days before the spacecraft was set to fire its main rocket engines and decelerate into orbit, flight controllers at NASA's Jet Propulsion lab (JPL) lost contact with the spacecraft—the mission failed. Similarly, the Space Shuttle program's paperwork began in earnest from September 1966. It was not formally kicked off until August 1968 when a request for proposal (RFP) was put to the market. Contracts were let out to General Dynamics, Lockheed, McDonnell Douglas, and North American Rockwell by December 1968. The first flight took place in April 1981. The Space Shuttle flew 135 times—with two fatal accidents, Challenger upon launch in 1986 and Columbia upon re-entry in 2003—until the programme's retirement in 2011. Therein lies the bigger issue with the Space Shuttle programme—conceived as a big one-off, bespoke project, the system could not be updated. A space vehicle designed in the 1960s and 1970s did not meet 21st Century standards and despite having spent $221 billion (in 2012 dollars) on the overall programme, the Space Shuttle had to be scrapped. The retirement of the Space Shuttle locked the US out of human space flight. It was only with the SpaceX Crew Dragon that this capability was recovered in May 2020.

NASA's Space Launch System (SLS)—a heavy lift rocket similar to SpaceX's Falcon Heavy or Starship—is an on-going case of a slow-to-market system. The SLS began life in 2011, around the same time as the Falcon Heavy, to replace NASA's Space

Atif Ansar & Bent Flyvbjerg 2022©                                                                    17

Shuttle. At the time of writing in December 2021, the first launch is not expected until February 2022. NASA has had to announce a delay in the SLS launch timetable at least half-a-dozen times since 2016 (Clark, 2021). The more substantive aim of launching humans into space—which SpaceX achieved three years ago—is not expected until 2024, assuming no further delays.

NASA is clearly slower to market and to iterate than SpaceX.

*Schedule overruns*
In terms of schedule overruns, although SpaceX is significantly quicker in absolute speed-to-market, it is just as optimistic about its schedules as NASA. Figure 4 shows optimism measured by schedule overruns of NASA missions compared against SpaceX. We observe:

Frequency
- 9 in 10 NASA missions suffered a schedule overrun. For SpaceX, the number is 10 out of 10.
- Size: For NASA the programme's actual duration of missions were on average +57.2% longer than estimated with a median of +34.8% again indicating that the distribution has a skew to the right (i.e., going over time). A Mann-Whitney U test of overall schedule neutrality supported this conclusion at an overwhelmingly high level of statistical significance ($p < 0.0001$), i.e., NASA's budgets are systematically biased towards underestimation. This results in a heavy bias towards adverse outcomes as shown in Figure 4.
- For SpaceX, the comparable numbers are an average schedule overrun of +39.8% with a median of +28.8%.
- Using a non-parametric Wilcoxon test, we tested whether the distribution of schedule overruns for NASA's space missions are different from SpaceX's. We found no significant difference ($p$ = 0.48), meaning that based on our sample SpaceX is neither more nor less optimistic than NASA in setting schedule targets. It should be mentioned, however, that the lack of significance could be due to the relative small size of the SpaceX sample. The test should be repeated once a larger sample exists.
- In practice, the +39.8% average schedule overrun implies that SpaceX promises delivery in three years but gets it done in four. Whereas for NASA, the +57.2% schedule overrun implies that NASA promises delivery in about four years but gets it done in six to seven.

Building on Flyvbjerg (2021c), we interpret SpaceX schedule slippage as a strategic bias more than a cognitive one. Elon Musk is known for making wildly optimistic schedule forecasts for when certain milestones will be achieved, not only for the SpaceX launches but also for Tesla car sales, factory openings, and the roll-out of charging stations (Vance, 2015, pp. 114-15, 143, 203, 308). This has been the case since Musk's very first company, Zip2, which developed an early version of Yellow Pages for the Internet.

Musk has explained why speed is so important to his ventures, and especially how costly long schedules and their overrun can be, using a six-month delay on the construction of Tesla's first Gigafactory, Giga Nevada, as an example, but the situation for SpaceX is similar: "Six months for this factory is a huge deal. Do the basic math and it's more than a billion dollars a month in lost revenue…Timing is important. We have to do everything we can to minimize the timing risk" (Vance, 2015, p. 329).

The pursuit of speed has led Musk to make strategically optimistic forecasts for how quickly something will get done to motivate his team and attract potential customers and investors. Counting delayed or lost revenues (or other benefits) as a cost is often overlooked in project planning and delivery, especially for public projects. With this focus, Musk decided long ago that he cannot afford long projects and therefore also cannot afford to use conventional design and delivery methods, which are notorious for being sluggish. Thus, he decided to reinvent design and delivery, whether for SpaceX, Tesla, or the construction of his gigafactories. For the US$5 billion Giga Nevada factory, he got to his revenue stream in less than a year, where this would typically take five years, following the conventional approach (Flyvbjerg 2021a).

**Figure 4: Density Trace and Mean (Vertical Lines) of Schedule Overruns: NASA (*N* = 122) vs SpaceX (*N* = 15)**

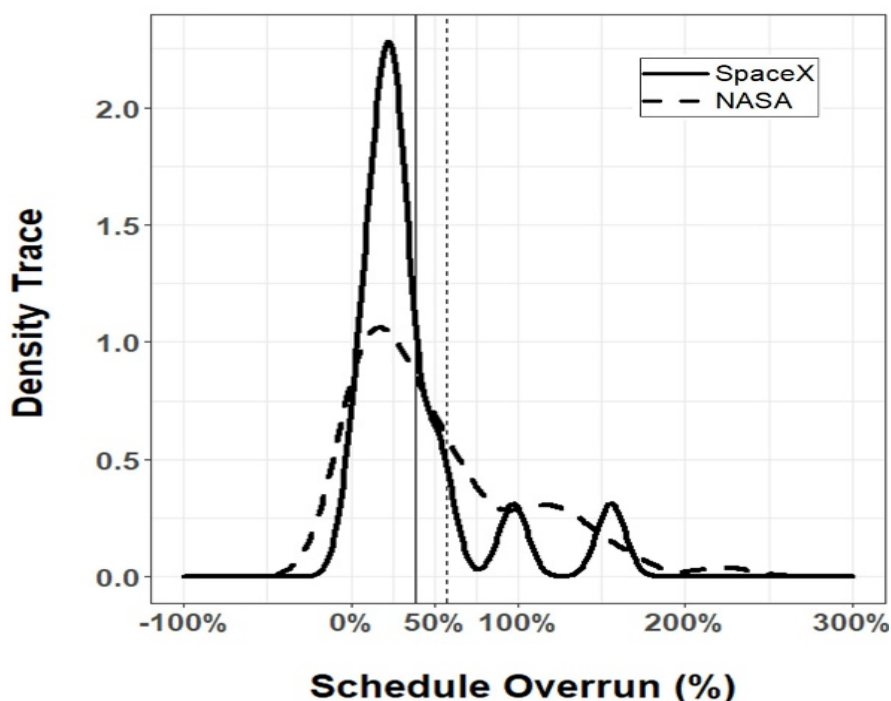

Source: Authors' Database

**Scalability**
We define scalability as the ability to handle an increased volume and variety of workload without crashing or needing a disproportionate number of additional resources. The concept is borrowed from computer science and is distinct from conventional notions of economies of scale in economics (see Ansar et al., 2016 for a discussion). To be truly scalable a system must be capable in handling both a surge of demand as it comes in and as the surge goes out. Google is an example of a highly scalable system. Millions of people searching for results on an election night does not crash Google. Similarly, once people lose interest in election results,

Google is not left stranded with investments in assets it cannot repurpose. It redeploys its algorithms, datacentres, and connectivity infrastructure to seamlessly address other searches. Scalable systems are fast, replicable, and adaptable in dealing with an increased volume and variety of workload (Ansar et al., 2016; Flyvbjerg, 2021).

As shown earlier, SpaceX is fast, NASA is slow. In highlighting the slow-moving pace of NASA's big one-offs, the intention here is not to undermine NASA's historic achievements. The Apollo programme that took humans to the Moon was a near-impossible feat achieved at a spectacular pace— seven years from Kennedy's announcement to Congress in May 1961 to the first flight in October 1968, with the first Moon landing a year later. But, as spectacular as it was, the Apollo programme was a bona fide one-off. There were six crewed US landings between 1969 and 1972 after which the programme ran out of money. The Apollo programme costs were unsustainably high (see 1961-1972 in Figure 5). There have been no human moon landings since, neither does the capability to enact one today exist. SpaceX is, in fact, the bet. Big one-offs do not enjoy rapidly declining cost or time-to-market curves. Platforms do.

The origin story of SpaceX has its source in Elon Musk's surprise, when as a newly minted millionaire with an interest in Mars he discovered that, despite spending billions of dollars annually over thirty years 1972-2002, NASA was nowhere near a human landing on Mars. In fact, NASA could not even return a few astronauts to the Moon, despite its enormous investments in the Moon flights. NASA's one-offs did not scale. SpaceX set out to build a platform that does.

**Figure 5: NASA's Annual Budget US$ Billion (2020 Constant Dollars)**

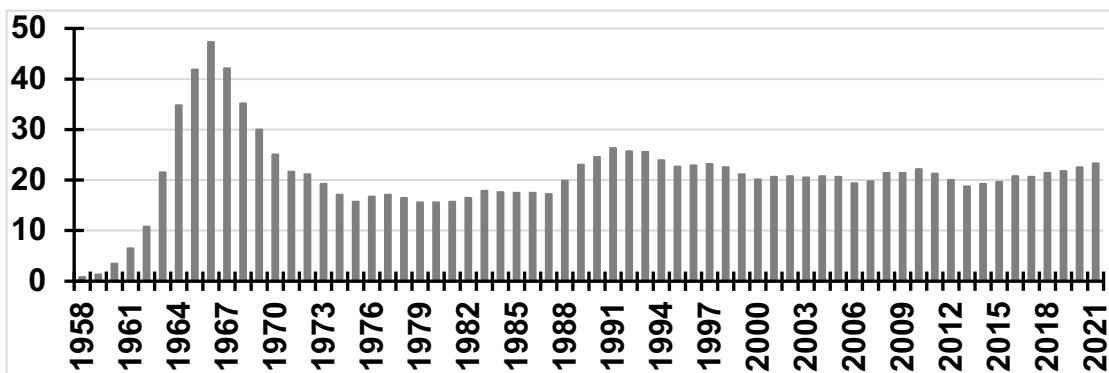

Source: NASA (Various Years) Budget Documents, Strategic Plans and Performance Reports: https://www.nasa.gov/budget

SpaceX's systems are replicable and adaptable. From the outset, Elon Musk made reusability of rockets, progressive iterations, and expandability essential pillars of SpaceX. "Musk reasoned that 'throwing away multimillion- dollar rocket stages after every flight makes no more sense than chucking away a 747 after every flight.' To Musk, reusability would be a key lever in generating commercial activity in the industry since, "the reason there is low demand for spaceflight is because it's ridiculously expensive…the problem is that rockets are not reusable" (Weinzierl et al. 2021, p. 4). In 2021, SpaceX landed a rocket for a 100$^{th}$ time, breaking a record in reusability (Zapata 2021). As seen earlier, SpaceX systems undergo rapid iterative upgrades. In terms of adaptability, successive SpaceX systems have absorbed the capabilities of the past systems and expanded the overall capability-set of subsequent generations. The SpaceX Starship, which was undergoing test flights in December 2021, will, for example, combine the reusability of Falcon 1 & 9, heavy-lift capability of Falcon Heavy, and human-rating of Crew Dragon system with the aim of carrying 100 people to Mars.

SpaceX is scaling fast in terms of volume and variety. In 2009, when SpaceX's future was still in doubt, the company's only commercial launch took RazakSAT—a Malaysian Earth Observation satellite weighing 180 kilograms—into orbit. Today, each of multiple annual launches can carry a multitude of crew and cargo into space, weighing up to 549,054 kilograms. In 2021, SpaceX set a record of 31 launches, with payloads ranging from military satellites to parts of its own megaconstellation of Starlink internet satellites. SpaceX has introduced the concept of "a rideshare" to the space market. For contract values as low as US$1 million, SpaceX bundles multiple types of space cargo and delivers it to the orbit. In June 2019, during Falcon Heavy rocket's first night launch, it sent 24 different spacecraft toward three different types of orbit. The payload included cargo as various as a privately funded solar sail experiment to harvest solar energy for interstellar flight, a NASA-designed miniaturized atomic clock for use in deep space, U.S. Pentagon-funded satellites to measure space radiation, and a container with the cremated remains of 152 people (Boyle, 2019).

In terms of volume, Figure 6 illustrates that in 2022, SpaceX's annual launches will tie with NASA's all-time high. NASA's budget exceeded US$40 billion in constant 2020 dollars in 1964, when NASA peaked with 42 launches. SpaceX's 2022 launch revenues are expected to be around US$2 billion for 41 launches. By 1970, NASA's budget had nearly halved and the launch frequency diminished. In contrast, SpaceX launches are on track to keep increasing in number, with its lower cost and higher speed-to-market. Edgar Zapata (2017b)—a Life Cycle Analyst at NASA for 32 years argues that 200+ annual launches—five times NASA's peak— are within reach for SpaceX.

**Figure 6: SpaceX is Scaling Fast**

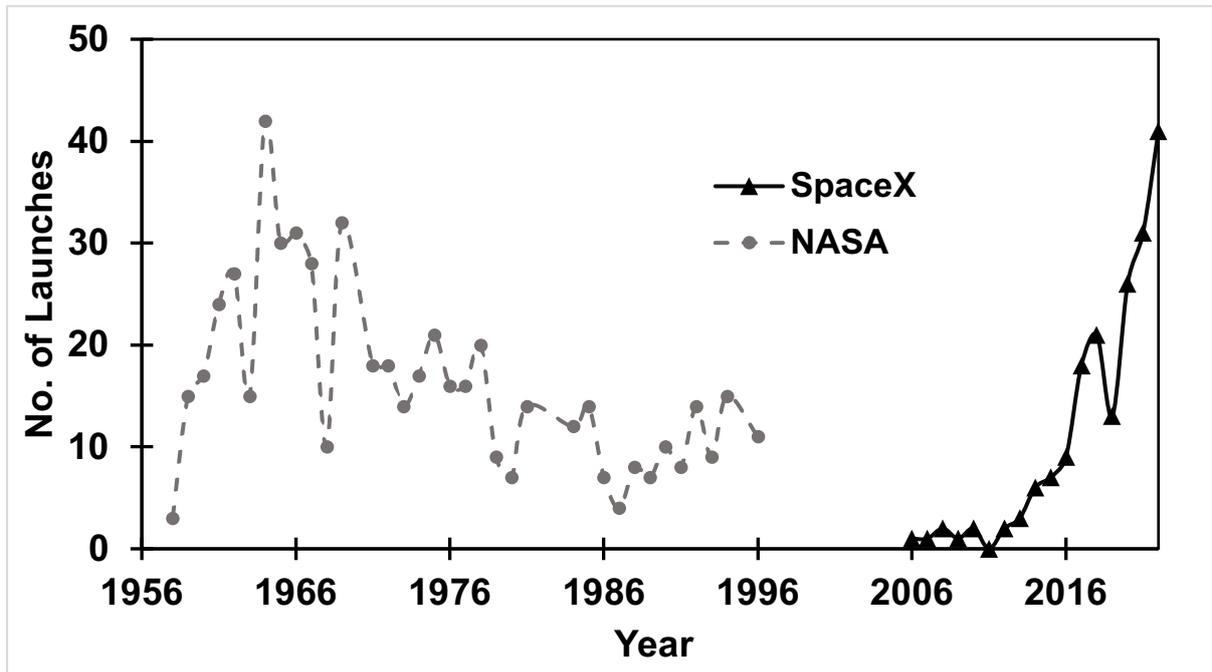

Source: Authors' Database

## DISCUSSION

How should government and business leaders solve big problems? By contrasting NASA's 20th Century quantum leap/bespoke strategy with SpaceX's 21st Century platform strategy we arrive at a set of general, if preliminary, lessons and policy propositions regarding how big problems are best solved: Not by one-off, bespoke solutions but by ones that are modular, iterate quickly, and scale up well. Specifically, we draw the following conclusions.

**Platforms Are Faster, Better, Cheaper, and Lower Risk**
Quantum-leap projects are not a reliable way to solve big problems like poverty, climate change, pandemics, prosperity, security, health and well-being. Our case study of the space industry indicates that a successful platform strategy is not only faster and cheaper, it also lowers risk. Figure 7 provides an impressionistic illustration of how moving from a bespoke strategy not only lowers the point estimate it also compresses the variance around the point estimates. That is, Space X missions are not only many times cheaper than conventional missions, they carry far less risk.

**Figure 7. Platforms Enable Lower Point Estimates and Lower Risk**

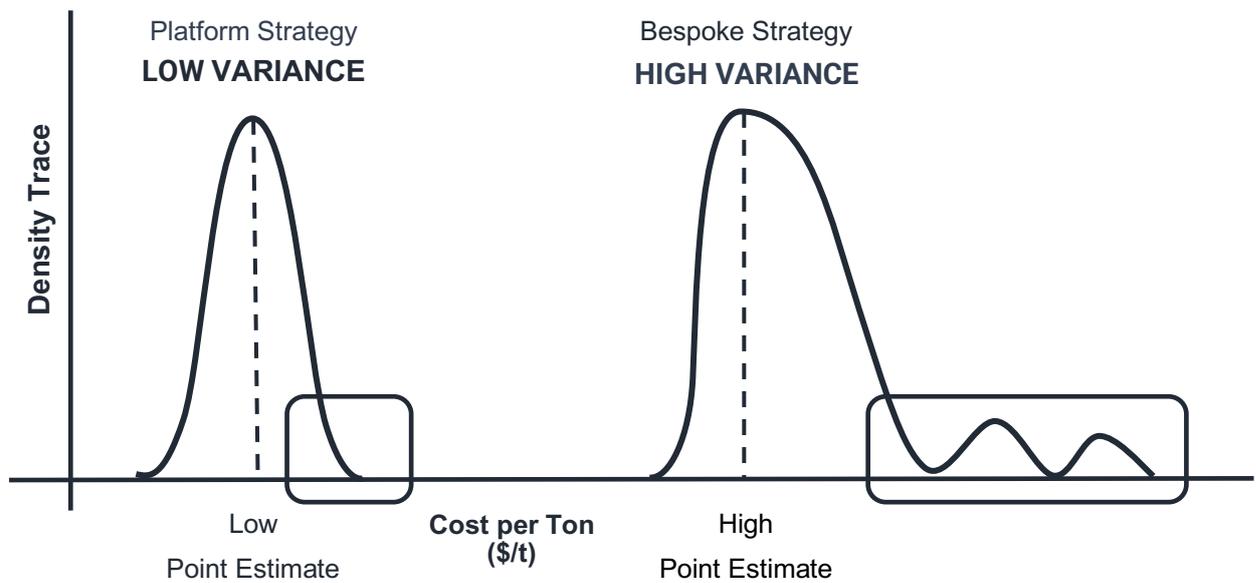

Source: Authors

Although here we focused narrowly on space missions, platforms are now widely understood in business practice and the management literature. Conventional economics and policy, however, have substantial catching up to do. Containerised shipping is 99% cheaper, 90% faster, and carries 30X the trading volume compared with the pre-platform era. Computing cycles are billions of times cheaper and faster. Platform companies such as Apple and Microsoft command multi-trillion-dollar valuations that exceed the GDP of the United Kingdom—the world's sixth-largest economy. Sectors of the economy where government finds it difficult to control spending or timeframes or to realize the benefits quickly enough – e.g., health, education, climate, defence – are ripe for a platform rethink.

Platforms defy the "Iron Triangle" myth. The "Iron Triangle" is sometimes also called "The Project Management Triangle" or the "Triple Constraint" (see Figure 8). The three corners of the Iron Triangle are made up of the variables scope, cost, and schedule. According to this model, the three variables are interrelated in a zero-sum game. Building a larger scope is thought to require more cost and time. Conversely, to build something faster is thought to either cost more or require smaller scope or both. The economics of platforms shows such conventional thinking to be mistaken. SpaceX delivers far greater scope—volume and variety of space cargo—than even its proponents had imagined. And it does it 10X cheaper and 2X faster than the conventional approach. Platforms defy the Iron Triangle. In fact, whether something can break the constraints of the Iron Triangle is a test for whether it will scale.

**Figure 8: The Iron Triangle Versus Platforms**

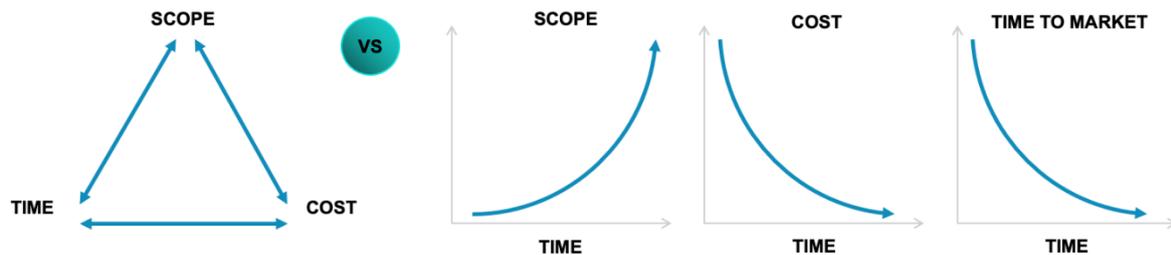

Source: Authors

The ability of well-executed platforms to defy the Iron Triangle presents a source for grounded optimism in the delivery of big projects. Not all big problems are as intractable as they appear.

> **Policy Proposition 1.** *Platform strategies outperform bespoke strategies. Governments and businesses facing large costs, lead times, and risk in solving big problems will improve performance by eschewing big one-off investments and switching to a platform strategy.*

**Platforms Enable Greater Volume *and* Variety**
The Volume Variety matrix is a useful tool to understand the power of platforms in solving big problems (Figure 9). Big problems typically require a scale (many units of outputs) and variety (many kinds of outputs). To populate outer space, as SpaceX aspires to do, is a big problem that will require a much larger volume of launches than is currently possible. Moreover, high-volume space ferrying will need to handle an increased variety of cargo: large numbers of people, tools and heavy equipment brought from Earth to space, resources mined in one part of space transported for use in another part etc.

In Figure 9, the horizontal axis describes the volume of outputs an organization produces. The vertical axis describes the variety of that output. The upper-left corner denotes one-off bespoke production—low volume and high variety with no standardization, e.g., works of art. The low-right corner denotes continuous production—high volume and low (or no) variety with high standardization, e.g., containerized shipping. Typically, a diagonal line is drawn between the bespoke and continuous ends of the matrix—as shown in Figure 8—denoting intermediate stages such as batch production or assembly line production. The unit cost of production declines as one travels down the diagonal.

**Figure 9: Volume/Variety Matrix**

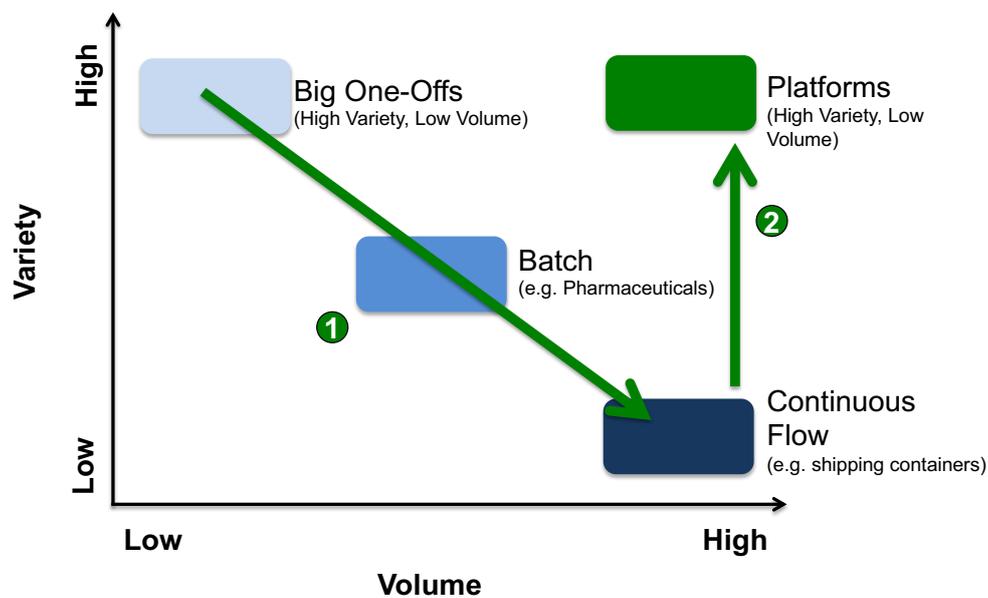

Source: Adapted from Hayes and Wheelwright (1979)

Conventional thinking stipulates a trade-off between volume and variety. NASA's one-off missions had a great deal of variety—the doomed Mars Observer carried 8 instruments for scientific discovery worth US$479 million (MacCormark & Wynn, 2004). Even if the Mars Observer had succeeded, its sequel would have taken just as much time and money to refashion each each element. In contrast, platforms are designed for repeatability through modularity. The initial advantage of this repeatability is to increase homogenous volume (diagonal line 1 in Figure 9). Once that repeatability is predictable, variety follows at scale by combining modules in different ways (vertical line 2 in Figure 9).

In a virtuous cycle, variety unlocks more volume. For example, SpaceX first achieved repeatable and reliable launches of its Falcon 9 rocket, which brought about significant cost reductions, which allowed lower prices, which spurred demand, e.g. from owners of small, inexpensive satellites. Then SpaceX began bundling payloads of different sizes to drive down costs further through the "rideshares" mentioned above.

Weinzierl et al. (2009) explain:

> These satellites were small enough (less than 500 kg) that once the majority of the Falcon 9 (with a capacity of 22,800 kg) had been filled with larger payloads, they could "piggyback" on a launch, filling up remaining space and enjoying low costs to orbit. Shortly after Falcon 9 became operational, the small satellite market saw a major boom, with a 23% compounded annual growth rate in 2009 to 2018. In 2017, these payloads made up 69% of total launches while only accounting for 4% of total mass. The greatest source of growth was the launch of large "constellations" or groups of small satellites. When operated in concert, small satellites achieved greater functionality, allowing for increased revisit times over areas on Earth or increased spatial resolution. Constellations for earth-imaging (such as Planet's Dove) and satellite broadband (such as SpaceX's Starlink) accounted for much of the demand in launch, as they included hundreds or even thousands of satellites.

Thus, as a platform grows, its ability to serve a greater scale and scope of services increases disproportionately. To illustrate, at its launch in June 2007, the Apple iPhone had three apps (phone, calendar, and music) and sold 1.39 million units. Today, over 2.5 billion units have been sold (volume) and the gadget has millions of apps and added features (variety).

> **Policy Proposition 2.** *Governments facing problems of increased volume and variety of services, budget cuts, and citizen pressure for speed-of-action will fail if they attempt to solve the problems with bespoke strategies. Platform-based policies that focus on producing a replicable core unit of output prior to step-wise scaling will outperform a quantum leap.*

**The Path to Platforms Is Through Failure, Iteration, and Learning**
Platforms exhibit increasing returns to scale. As a platform grows larger, its value increases disproportionately. At its fifth launch, SpaceX's Falcon 1 rocket could only deliver one small Earth Observation satellite and the rocket then had to be retired. A decade later, with the successful launch and landing of Starship SN15, platform was close to returning humans to the Moon (in the hundreds) and make the first human landing on Mars. The path to these increasing returns to scale paradoxically do not come from economies of scale. Instead, they come from economies of learning.

Organizations learn via multiple processes, for example, learning-by-doing, trial-and-error learning, experimental learning, vicarious learning, and improvisational learning, explained in more detail below (Bingham and Davis, 2012). Although someone within an organization learns something new every day, at the aggregate organizational level that learning can become disjointed or myopic (Levinthal and March, 1993; Lane et al. 2006). Even if learning embeds at the organizational level, there remains the problem of exploiting the learning (March, 1991). Consequently, unless an organization is radically disciplined, the default is a low learning adsorptive capacity and commensurately low exploitation of obtainable learning to achieve organizational purposes.

Platforms codify learning in a disciplined manner, what Bingham and Davis (2012) call "learning sequences". Research has shown that learning sequences exist, that they matter to both short and long-term performance, and that managers can orchestrate such sequences for more effective learning. SpaceX began its orderly learning sequences with experimental learning. SpaceX also deliberately avoided vicarious learning, i.e., learning by mimicking others in the industry (Ingram, 2002). If at SpaceX someone would say, "This is how we always did it at NASA", it would likely get them fired (Berger, 2021).

Experimental learning takes place in controlled "offline" situations, the proverbial laboratory environment, that organizations use to test hypotheses (see Miner et al. 2001, Table 2, p. 319). As the chief purpose of experimentation is to confirm or disconfirm hypothesized causal relationships, ex post reflection on outcomes and data analysis is high in disciplined firms.  Experimental learning is different from learning-by-doing or trial-and-error learning: "in experimental learning, variation in conditions is planned and intentionally introduced to produce insights about input-output relations" typically about unknowns (Bingham and David, 2012, p. 613). Trial-and-error, in contrast takes place in "on-line" environments typically to refine ongoing knowns or regular activities.

From the outset, the big goal for Space X has been to enable a multi-planet society and specifically to populate Mars. The company started with a blank sheet in 2002 with a founder who at the time admitted to no knowing much about the physics and engineering of space travel, but knew enough to know he did not want to do things the NASA way. Instead he wanted to rethink space travel from basic principles. SpaceX thus had several fundamental hypotheses to test: Could it build a rocket engine? Could it fire up the engine without the engine blowing up? Could the engine power a rocket to go vertical? Could the rocket deliver a payload—no matter how small the cargo—to space? Could the company reuse the rocket? Could it land back a rocket it had flown? Could it do all this at a cost10 times lower than rivals? Could it attract paying customers? Could it generate revenue from non-public sector customers? Could it remain solvent? Could it get regulatory approvals to launch? Could it get a launch base? Could it get human certification? Could its rockets carry human crews as reliably as a 747? Could it increase the payload 10X? 100X? None of these questions had clear answers at the outset.

Figure 10 shows SpaceX's learning sequence from 2002 to 2022 on its quest to land humans on Mars has taken the form of a "learning staircase"—deliberate incremental moves of increasing stakes that, as they have combined over time, have generated disproportionately larger returns. The outcome of an experiment cannot be known in advance, otherwise it is not an experiment. SpaceX has thus experienced, and sought, failure at each step (see Seedhouse, 2013; Berger, 2021). Failure in such experimental sequences can be defined in two ways—where the experiment disconfirms the initial hypothesis. The failed experiment thus detects that the experimenter's underlying theory has an anomaly that needs revision. In other instances, where the experiment supports the initial hypothesis but something else breaks down, reveals a blind spot that needs correction in future iterations.

Flyvbjerg (2021a) calls such sequences "positive learning curves," whereby a task at hand gets faster, cheaper, and less effortful as it is performed more and more. Counterintuitively, however, some things, the more one does them the more difficult, slow, and expensive they become. This is called "negative learning curves." In extreme cases, negative learning may result in aborted ventures. NASA's Space Shuttle had negative learning curves and was aborted. SpaceX's rockets have positive learning curves and have rapidly reinforced progress towards Space X's mission and increasing the value of the company.

Each step of SpaceX's learning staircase thus represents validated learning that was consolidated through a process of prototyping, experimental testing, failing, revising, forging forwards. Other forms of learning, such as improvisational bricolage to solve surprises or take advantage of unexpected opportunities converged with this experimentation-led learning sequence (Miner et al., 2001; Bechky and Okhuysen, 2011).  Unlike SpaceX's multi-step learning sequence, the Mars Observer mission attempted a single-step quantum leap over a 17-year period—it failed. Lacking a clear mission-orientation, NASA's internal Mars programme remains a hodgepodge of initiatives. If humans are to land on Mars in the coming decade, it will be aboard a SpaceX vessel.

**Policy Proposition 3.** Policies that pursue a big goal with a disciplined sequence of experiments of increasingly higher stakes will outperform single-shot solutions.

**Figure 10: Platform Learning Staircase: Step-Wise Scaling of SpaceX (2002-2022)**

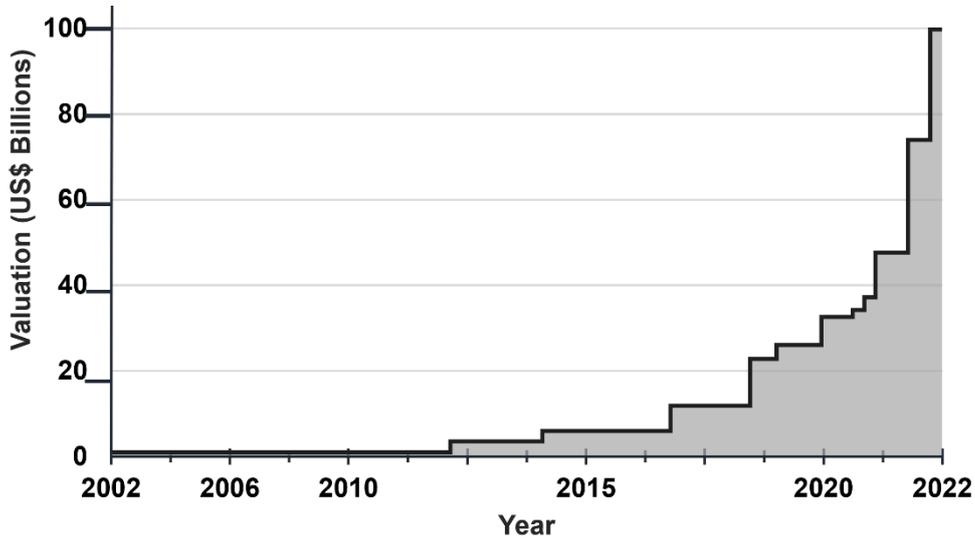

| 2002 | SpaceX founded |
| 2006 | First Falcon 1 launch fails |
| 2007 | Second Falcon 1 launch fails |
| 2008 | Third Falcon 1 launch fails |
| 2008 | Falcon 1 Flight 4 succeeds and reaches orbit |
| 2010 | Dragon spacecraft returns to Earth |
| 2012 | Cargo flights to ISS begin |
| 2015 | Falcon 9 explodes after launch |
| 2016 | Falcon 9 explodes before launch during fuelling |
| 2016 | First successful landing of Falcon 9 first stage and first water landing |
| 2017 | Successful recovery and reuse of rockets, global first reflight of an orbital class rocket |
| 2018 | Falcon Heavy first launch |
| 2019 | First batch of Starlink satellites launched |
| 2020 | First crewed mission to ISS |
| 2020 | 100th SpaceX launch |
| 2021 | Starship SN11 rocket crashes during landing in March |
| 2021 | Starship SN15 successfully launches and lands in May |

Source: Timeline of SpaceX https://timelines.issarice.com/wiki/Timeline_of_SpaceX

The practice of government policymaking is mired in failure avoidance—colloquially called bureaucracy. Because experiments have rapid feedback loops, failures and successes are discovered quickly. The apparatus of contemporary government—with a few inevitable exceptions—is unable to handle the multiple micro failures experiments yield, essential to macro success. Instead, governments opt for macro failure with big, slow quantum leaps, where, the failure can be smudged with the fog of intervening time. NASA, to its credit, was on to platform thinking with NASA Administrator Dan Goldin's push for "Faster, Better, Cheaper (FBC)" iterations as early as 1992 (see McCurdy, 2001, pp. 50-51)[3]. However, failures of just two programs launched under the FBC unnerved the organization and NASA reverted to its old ways. Fortunately, NASA had the foresight and courage to create the COTS/CRS program in the mid-2000s that protected an experimental learning approach from the choke of bureaucracy. If and when humans land on Mars aboard a SpaceX vessel, vicariously the achievement will also be NASA's.

> **Policy Proposition 4.** Governments averse to the micro-failures of experimental learning will default to taking much larger failure risk on single-shot solutions. For government to be effective, policymakers will need to muster the courage to set up experimental learning sequences at arms-length of the incumbent institution.
>
> **Policy Proposition 5.** Any government or business that aims to successfully deliver its products and services at scale will need to use a platform strategy singularly focused on positive learning (modular, fast, iterative) and stay away from the negative learning that is characteristic of quantum-leap strategies (bespoke, slow, one-off).

**CONCLUSIONS**

The empirical evidence presented in this paper contributes to discussions of how government and business should solve big problems. It demonstrates that big, bespoke projects are not a reliable way to solve problems. Rather, governments and businesses should adopt platform strategies like that of SpaceX. By fundamentally challenging conventional thinking about major projects and their inherent risk, platform strategies have made possible feats that only decades ago would have been unimaginable. Not only this, but these achievements have been carried off in record time. While the thrust of our comparative case study rests on the achievements of SpaceX versus NASA, clear evidence exists across multiple sectors -- containerised shipping, computing, car manufacturing, solar energy, wind power, batteries --  to demonstrate that the platform approach has already enacted sea-changes in society.

On the basis of this evidence, it is advised that a platform approach be considered for large-scale problem-solving ventures of government (as outlined in the Policy Propositions above). Looking to the future, a platform approach can ensure that failures like the National Programme for Information Technology (NPfIT) become a thing of the past. In the context of the climate crisis and increasing levels of political uncertainty, it has become more imperative than ever that governments manage the myriad of challenges they face with minimised risk and high levels of adaptability to change. The answer to this is a platform approach.

---

[3] McCurdy (2001, pp. 50-51) quotes Dan Goldin speech in the Autumn of 1992, "We should send a series of small and medium-sized robotic spacecraft to all the planets and major moons, as well as some asteroids and comets. Let's see how many we can build that weigh hundreds, not thousands, of pounds; that use cutting-edge technology, not 10-year-old technology that plays it safe; that cost tens and hundreds of millions, not billions; and take months and years, not decades, to build and arrive at their destination. Slice through the Gordian knot of big, expensive spacecraft that take forever to finish. By building them assembly line style, we can launch lots of them, so if we lose a few due to the riskier nature of high technology, it won't be the scientific disaster or blow to national prestige that it is when you pile everything on one probe and launch it every ten years."

Government Accountability Office. (2011). *NASA: Assessments of Selected Large-Scale Projects: GAO-11-239SP.* http://www.gao.gov/assets/320/316257.pdf

Government Accountability Office. (2013). *NASA: Assessments of Selected Large-Scale Projects: GAO-13-276SP.* https://www.gao.gov/products/gao-13-276sp

Government Accountability Office. (2014). *NASA: Assessments of Selected Large-Scale Projects: GAO-14-338SP.* https://www.gao.gov/products/GAO-14-338SP

Government Accountability Office. (2015). *NASA: Assessments of Selected Large-Scale Projects: GAO-15-320SP.* http://www.gao.gov/assets/670/669205.pdf

Government Accountability Office. (2016). *NASA: Assessments of Major Projects: GAO-16-309SP* (GAO-16-309SP). U.S. Government Accountability Office. https://www.gao.gov/products/gao-16-309sp

Government Accountability Office. (2017). *NASA: Assessments of Major Projects: GAO-17-303SP.* https://www.gao.gov/products/gao-17-303sp

Hagiu, A., & Wright, J. (2015). *Multi-Side Platform* (No. 15–037). Harvard Business School.

*Harvard Business Publishing Education*. (n.d.). Retrieved 2 January 2022, from https://hbsp.harvard.edu/download?url=%2Fcatalog%2Fsample%2F909E09-PDF-ENG%2Fcontent&metadata=e30%3D

Hayes, R. H., & Garvin, D. A. (1982). Managing as If Tomorrow Mattered. *Harvard Business Review*, *60*(3), 70–79.

Hayes, R. H., & Wheelwright, S. C. (1979, February 1). Link Manufacturing Process and Product Life Cycles. *Harvard Business Review*, *57*(1), 133–140.

Helm, D., & Tindall, T. (2009). The Evolution of Infrastructure and Utility Ownership and Its Implications. *Oxford Review of Economic Policy*, *25*(3), 411–434. https://doi.org/10.1093/oxrep/grp025

Hendy, J., Reeves, B. C., Fulop, N., Hutchings, A., & Masseria, C. (2005). Challenges to implementing the national programme for information technology (NPfIT): A qualitative study. *BMJ*, *331*(7512), 331–336. https://doi.org/10.1136/bmj.331.7512.331

Hirschman, A. O. (1967). *Development Projects Observed*. Brookings Institution Press; JSTOR. https://www.jstor.org/stable/10.7864/j.ctt7zsw04

Ingram, P. (n.d.). Interorganizational Learning. In J. A. C. Baum (Ed.), *Companion to organizations* (pp. 642–663). Blackwell.

Jones, H. (2018). *The Recent Large Reduction in Space Launch Cost*. https://ttu-ir.tdl.org/handle/2346/74082

Kahneman, D., & Tversky, A. (1979). Intuitive Prediction: Biases and Corrective Procedures. In S. Makridakis & S. C. Wheelwright (Eds.), *A Discounting Framework for Choice With Delayed and Probabilistic Rewards*. North Holland.

King, A., & Crewe, I. (2013). *The Blunders of Our Governments*. Oneworld Publications.

Krygier, M. (2012). *Philip Selznick: Ideals in the World*. Stanford Law Books. http://ezproxy-prd.bodleian.ox.ac.uk:2317/view/10.11126/stanford/9780804744751.001.0001/upso-9780804744751

Kwoka, J. E. (2002). Vertical Economies in Electric Power: Evidence on Integration and Its Alternatives. *International Journal of Industrial Organization*, *20*(5), 653–671.

Lambrecht, B. M. (2004). The Timing and Terms of Mergers Motivated by Economies of Scale. *Journal of Financial Economics*, *72*(1), 41–62.

Lane, P. J., Koka, B. R., & Pathak, S. (2006). The Reification of Absorptive Capacity: A Critical Review and Rejuvenation of the Construct. *Academy of Management Review*, *31*(4), 833–863.

Levinson, M. (2006). *The Box: How the Shipping Container Made the World Smaller and the World Economy Bigger*. Princeton University Press.

Levinthal, D. A., & March, J. G. (1993). The Myopia of Learning. *Strategic Management Journal*, *14*(S2), 95–112.

Lindblom, C. E. (1959). The Science of" Muddling Through". *Public Administration Review*, 79–88.

Lovallo, D., Clarke, C., & Camerer, C. (2012). Robust Analogizing and the Outside View: Two Empirical Tests of Case-Based Decision Making. *Strategic Management Journal*, *33*(5), 496–512. https://doi.org/10.1002/smj.962

MacCormack, A., & Wynn, J. (2004). *Missioin to Mars* (No. 9-603–083). Harvard Business School.

Majumdar, S. K., & Venkataraman, S. (1998). Network Effects and the Adoption of New Technology: Evidence from the U. S. Telecommunications Industry. *Strategic Management Journal*, *19*(11), 1045–1062.

March, J. G. (1991). Exploration and Exploitation in Organizational Learning. *Organization Science*, *2*(1), 71–87.